\documentclass[a4paper]{jpconf}
\usepackage{graphicx}
\usepackage{hyperref}
\usepackage{xspace}

\newcommand{\gambit}{\textsf{GAMBIT}\xspace}

\newcommand{\darkbit}{\textsf{DarkBit}\xspace}

\newcommand{\GB}{\gambit}

\newcommand{\ds}{\textsf{DarkSUSY}\xspace}
\newcommand{\darksusy}{\ds}

\newcommand{\micromegas}{\textsf{micrOMEGAs}\xspace}

\newcommand\nulike{\textsf{nulike}\xspace}
\newcommand\gamLike{\textsf{gamLike}\xspace}
\newcommand\gamlike{\gamLike}

\newcommand\ddcalc{\textsf{DDCalc}\xspace}

\newcommand\calchep{\textsf{CalcHEP}\xspace}

\begin{document}
\title{An overview of DarkBit, the GAMBIT dark matter module}

\author{Jonathan M. Cornell, on behalf of the GAMBIT collaboration}

\address{Department of Physics, McGill University,
3600 Rue University, Montr\'eal, Qu\'ebec, Canada H3A 2T8}

\ead{cornellj@physics.mcgill.ca}

\begin{abstract}
In this conference paper, I give an overview of the capabilities of DarkBit, a module of the GAMBIT global fitting code that calculates a range of dark matter observables and corresponding experimental likelihood functions. Included in the code are limits from the dark matter relic density, multiple direct detection experiments, and indirect searches in gamma-rays and neutrinos. I discuss the capabilities of the code, and then present recent results of GAMBIT scans of the parameter space of the minimal supersymmetric standard model, with a focus on sensitivities of future dark matter searches to the current best fit regions.
\end{abstract}

\section{Introduction}

The Global and Modular Beyond-the-Standard-Model (BSM) Inference Tool (\gambit) (\mbox{\url{http://gambit.hepforge.org}}) \cite{Athron:2017ard} is a recently released public code that serves as a flexible framework for fits of BSM theory parameters to a wide range of experimental constraints. The code has the ability to calculate limits from many probes of new physics, including production of new particles at colliders \cite{Balazs:2017moi}, flavour physics \cite{Workgroup:2017myk}, precision observables such as the anomalous magnetic moment of the muon \cite{Workgroup:2017bkh} and dark matter (DM) searches \cite{Workgroup:2017lvb}. In this conference paper, I will focus on the latter and give an overview of the DM observables and likelihoods that the code currently contains. I will also show results of scans of the minimal supersymmetric standard model (MSSM) \cite{Athron:2017yua} and forecasts of the ability of future DM searches to probe the best fit regions identified in these scans.

\section{Relic density}
\begin{figure}
  \centering
  \includegraphics[width=0.49\textwidth]{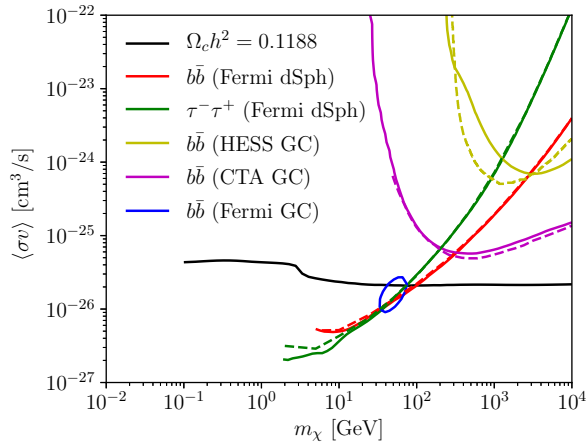} \\
   \caption{95\% CL limits on the DM annihilation cross section  from searches for DM annihilation in dwarf spheroidal galaxies with the Fermi-LAT \cite{Ackermann:2015zua} and in the galactic centre with H.E.S.S. \cite{Abramowski:2011hc}. Also shown are projected limits from CTA \cite{Silverwood:2014yza} observations of the galactic halo and the 99.7\% CL best fit region for the galactic centre excess, as determined in \cite{Calore:2014xka}. 
For all limits, the limit from \gamlike is shown as a solid line, while the corresponding official limits are dashed  (the discrepancies in the H.E.S.S. results are due to different adopted photon yields). Finally, the solid black line shows \darkbit's calculation of the values of the annihilation cross section that give $\Omega_c h^2 = 0.1188$ \cite{Ade:2015xua}.}.
  \label{Fig:ann}
\end{figure}

One of the most important aspects of any DM model is that it has a mechanism via which the observed abundance of DM in the universe can be produced. For WIMP-like models, \darkbit calculates the relic density via its interfaces to two well known external packages, \darksusy \cite{darksusy} and \micromegas \cite{Belanger:2013oya}. While \micromegas for some time has been able to calculate the relic density for an arbitrary model via its interface with the \calchep matrix element generator, a similar calculation has been more challenging with \darksusy, which has been designed with a focus on calculations in the MSSM, only one of the models we plan to ultimately implement in \GB. The modular nature of \GB makes it easy to give the \darksusy Boltzmann equation solver an arbitrary function for the invariant annihlation rate, enabling \darksusy to calculate the relic density for a non-SUSY model. To demonstrate this, we have used \darksusy to calculate the relic density for a generic WIMP model; the values of the $s$-wave annihilation cross section needed to obtain the correct relic density as determined from \textit{Planck} observations of the cosmic microwave background \cite{Ade:2015xua} are shown in Fig.~\ref{Fig:ann}. We have also used \darksusy to determine the relic density in our scan of the scalar singlet DM model \cite{Athron:2017kgt} as well as our scans of the MSSM \cite{Athron:2017qdc,Athron:2017yua}.

\section{Direct detection}
\begin{figure}
  \centering
  \includegraphics[width=0.49\textwidth]{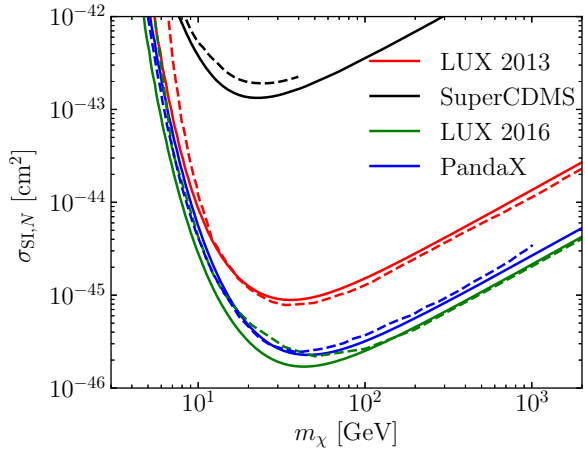} \hspace{2pc}%
  \begin{minipage}[b]{14pc}\caption{90\% CL limits on the DM-nucleon spin-independent scattering cross section from the SuperCDMS \cite{SuperCDMS}, LUX \cite{LUX2013,LUXrun2}, and PandaX \cite{PandaX2016} experiments as determined by DDCalc (solid lines), and the corresponding official 90\% CL limits reported by the experimental collaborations (dashed lines).}
  \label{Fig:DDCalc}
  \end{minipage}
\end{figure}

Direct detection (DD) experiments usually only present limits on the DM-nucleon scattering cross section for simple models in which DM interactions with either protons or neutrons can be described by a solely spin-independent or spin-dependent scattering cross section. While these simple scenarios map well to many models of DM in the literature, it is also possible for a DM candidate to have a mix of different types of interactions with nucleons (\textit{e.g.} in the MSSM). Furthermore, the usefulness of these results is limited by the fact that the limits are solely given at 90\% CL, and for a global fit 
we need a likelihood function for the entire parameter space. In parallel with the development of GAMBIT, we have developed a new standalone tool, \ddcalc (\url{http://ddcalc.hepforge.org}), that calculates a likelihood for an arbitrary DM model based on the number of events observed and expected backgrounds reported by the experiments. \ddcalc contains multiple experimental results, including the most recent constraints from Xenon1T \cite{Xenon1T2017}, PICO-60 \cite{PICO2017} (both newly added to version 1.1.0), PandaX \cite{PandaX2016}, and LUX \cite{LUXrun2}. Fig.~\ref{Fig:DDCalc} shows how the limits on the scattering cross section determined by \ddcalc vary by at most a factor of 1.5 from those reported by the experimental collaborations.

\section{Indirect detection}

\subsection{Gamma rays}
To determine the spectrum of gamma rays from a single DM annihilation, \darkbit makes use of tabulated spectra from either \darksusy or \micromegas if the annihilation is to a purely standard model (SM) final state. If the annihilation is to a new particle which then cascade decays down to a SM particle, \darkbit uses a newly developed fast cascade Monte Carlo code to simulate the cascades on the fly and determine the ultimate gamma-ray spectrum.

The likelihood functions for searches for DM in gamma rays are calculated using the \gamlike (\url{http://gamlike.hepforge.org}) code, released in parallel with \gambit. Using tabulated gamma-ray yields from a single DM annihilation as well as the DM annihilation cross section, \gamlike calculates the expected flux and resulting likelihood for gamma-ray searches in a variety of astrophysical targets. These include a stacked analysis of dwarf spheroidal galaxies based on likelihoods released by the \textit{Fermi} Collaboration \cite{Ackermann:2015zua}, a fit to the galactic centre excess as extracted from Fermi-LAT data in \cite{Calore:2014xka}, limits from observations of the galactic centre with H.E.S.S. \cite{Abramowski:2011hc}, and projected sensitivities from CTA observations of the galactic halo \cite{Silverwood:2014yza}.

\subsection{Solar Neutrinos}
Since limits from neutrino telescopes on the annihilation rate of DM are weaker than those from observations of gamma rays, we have chosen to focus on searches for high energy neutrinos from DM annihilation in the sun. These searches are used to constrain the DM-nucleon scattering cross section which enters into the DM capture rate. \darksusy is used to determine the expected flux of neutrinos at Earth, and then the publicly available code \nulike \cite{Aartsen:2016exj} is used calculate the likelihood of a model based on event-level energy and angular information from the IceCube telescope. Both 22-string \cite{Abbasi:2009uz} and 79-string \cite{Aartsen:2016exj} analyses are currently available.

\section {MSSM scan results}

We have used \gambit to carry out an extensive scan of a seven parameter version of the MSSM. This model (which is described in more detail in \cite{Athron:2017yua}) is defined in terms of  unified gaugino $M_2$ and sfermion $m^2_{\tilde f}$ mass parameters, the Higgs sector parameters $M^2_{H_u}$, $M^2_{H_d}$, and $\tan \beta$, and two third-generation trilinear couplings $A_{u_3}$, and $A_{d_3}$. We also varied multiple nuisance parameters, namely the strong coupling, top quark pole mass, local DM density, and nuclear matrix elements relevant for spin-independent DD. A range of collider, flavour, and precision physics likelihoods were included in the composite likelihood, as well as all the DM observables described above. The relic density was constrained by the condition that the calculated value not exceed the measured value $\Omega_c$ \cite{Ade:2015xua}, and we rescaled the DM density by a factor $f = \Omega_{\tilde \chi^0_1} / \Omega_c$, where $\Omega_{\tilde \chi^0_1}$ is the calculated neutralino relic density. This factor reflects the possibility that only the neutralino portion of the total DM density contributes to signals in direct and indirect detection experiments \cite{Bertone:2010rv}.

\begin{figure}
  \centering
  \includegraphics[width=0.49\textwidth]{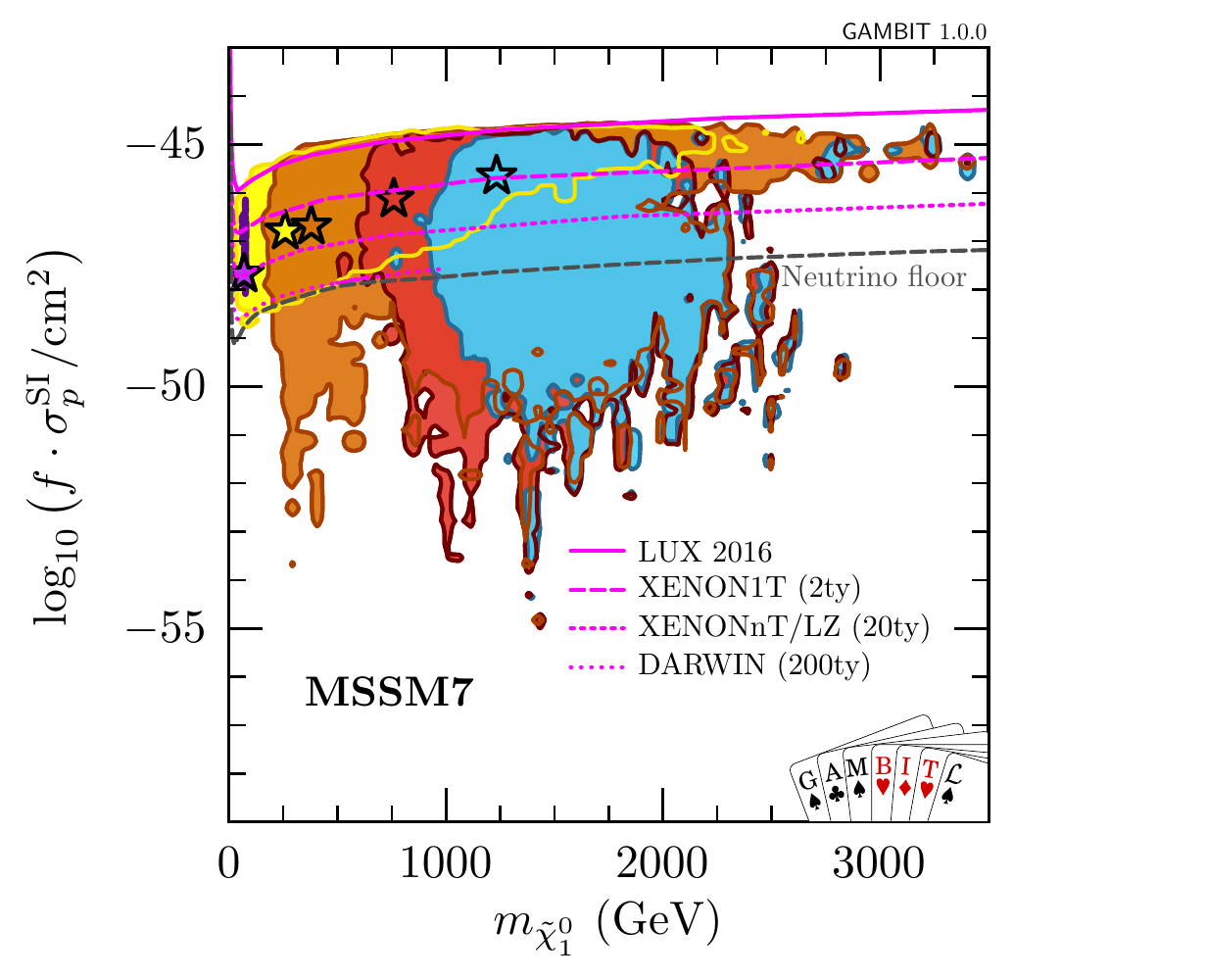}%
  \includegraphics[width=0.49\textwidth]{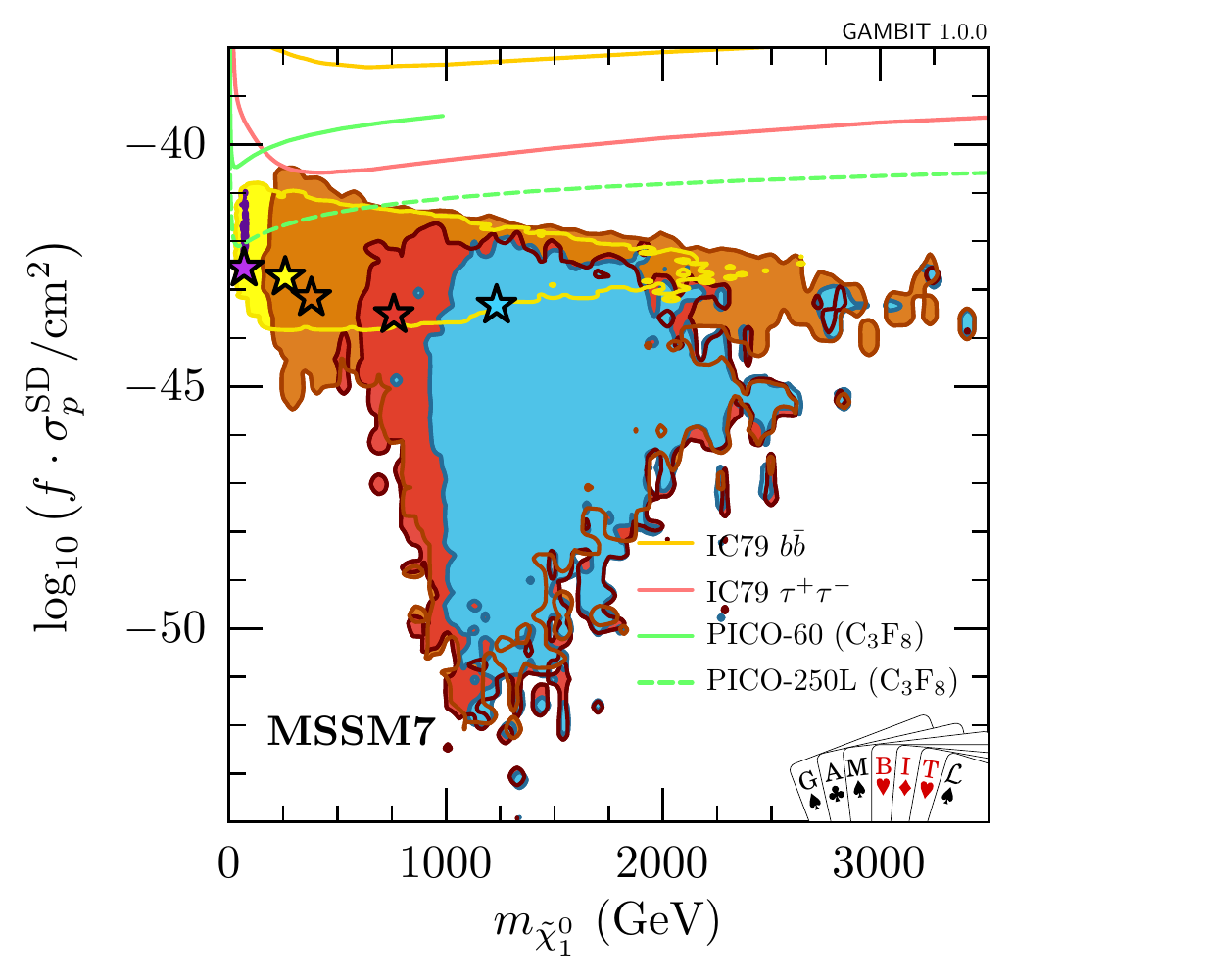}\\
  \includegraphics[width=0.99\textwidth]{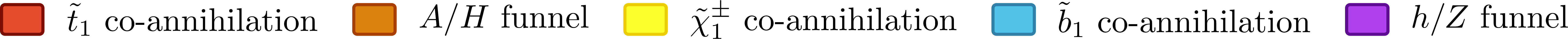}
  \caption{Regions of neutralino-nucleon scattering cross section vs. neutralino mass that are preferred at 95\% CL (all other parameters are profiled over). The left hand plot shows spin-independent cross sections, while the right displays results for the spin-dependent case. In both cases, the cross section is rescaled by $f$, a factor that compensates for the reduced event rate when the calculated relic density is less than the measured value. The regions are colour-coded by the processes which predominantly contribute to the neutralino annihilation rate in the early universe. The non-solid lines correspond to the expected sensitivities of future DD searches \cite{XENONnTLZ,DARWIN,Amole:2015cca}, and the stars represent the best fit points for each relic density mechanism.}
  \label{DD}
\end{figure}

In Fig.~\ref{DD}, plots showing the neutralino-nucleon scattering cross sections and neutralino masses that our scan has identified as being preferred at 95\% CL are displayed. To demonstrate the ability of future direct searches to test the MSSM, we have also plotted curves displaying the expected eventual sensitivities of these experiments. Ultimately, it is spin-independent DD searches which have the greatest capability to explore the preferred MSSM parameter space; proposed searches by the XENONnT/LZ \cite{XENONnTLZ} and Darwin  \cite{DARWIN} collaborations should probe nearly all of the 68\% CL preferred regions, largely excluding the chargino co-annihilation region, and will also explore much of the region preferred at 95\% CL. For spin-dependent DD, even the largest proposed version of the PICO experiment, PICO-250 \cite{Amole:2015cca}, will only probe a small region of the viable parameter space. 

\begin{figure}  
  \centering
  \includegraphics[width=0.49\textwidth]{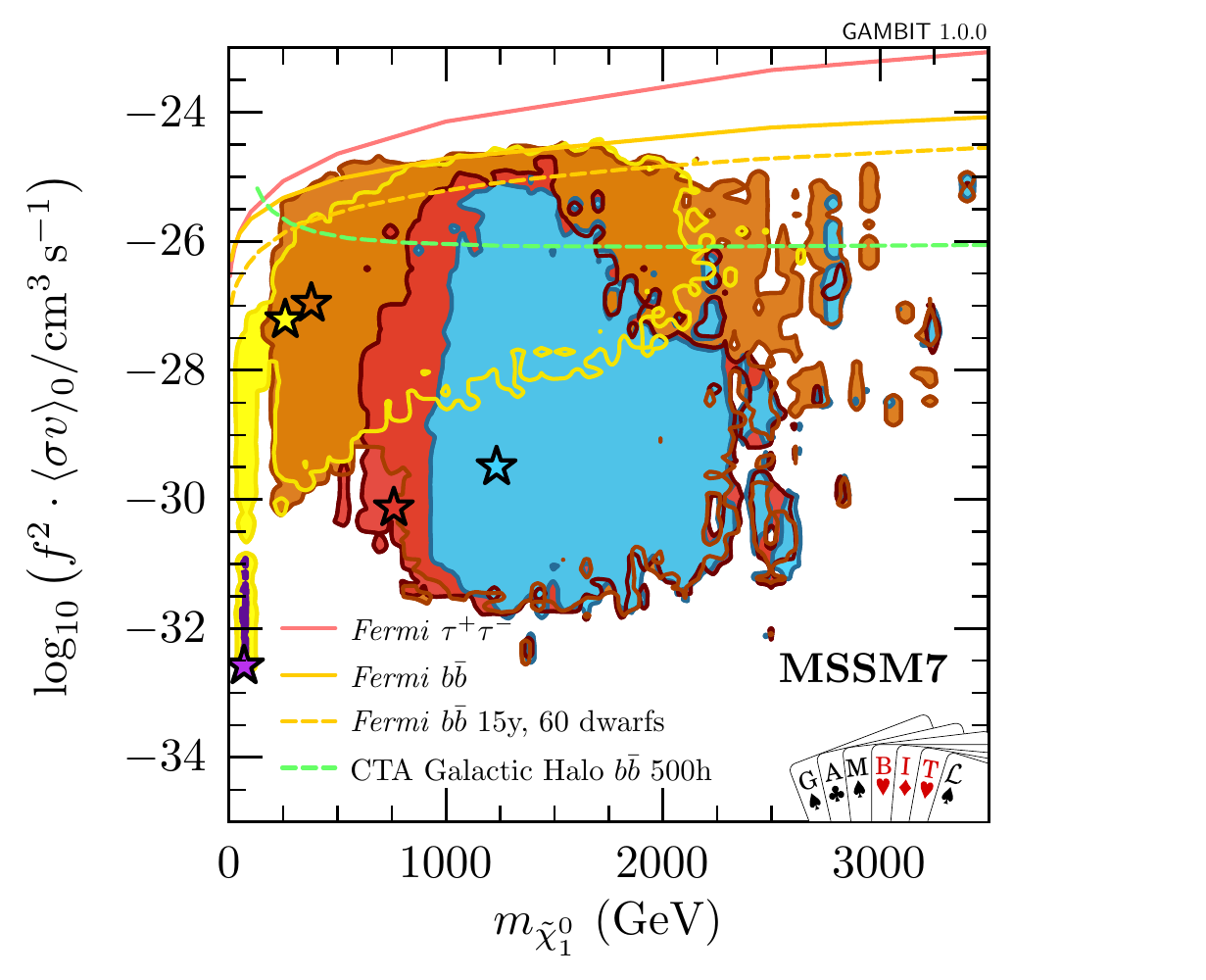}\hspace{2pc}
  \begin{minipage}[b]{14pc}\caption {Regions of neutralino self-annihilation cross section vs. neutralino mass that are preferred at 95\% CL. $f$, the stars, and the colour coding are as in Fig.~\ref{DD}. The lines correspond to current and future projected limits on the DM self-annihilation cross section from Fermi-LAT searches in dwarf spheroidal galaxies \cite{Ackermann:2015zua,Charles:2016pgz} and CTA observations of the galactic centre \cite{Carr:2015hta}.}
  \label{ID}
  \end{minipage}
\end{figure}

Fig.~\ref{ID} is a a similar plot showing the range of neutralino self-annihilation cross sections that are currently preferred, with projected sensitivities from future gamma-ray searches for DM annihilation overlaid. CTA observations of the galactic centre \cite{Carr:2015hta} have the ability to explore much of this parameter space, but there is still a large region which is below the sensitivity of future gamma-ray observatories. The models in this region have substantially enhanced neutralino annihilation rates in the early universe due to either resonance effects or co-annihilations, leading to small self-annihilation cross sections today.

\section{Summary and future plans}
\darkbit is a tool for the calculation of the likelihoods for arbitrary models of DM from a range of experimental DM searches. It is now publicly available and can be used both as part of the larger \gambit global fitting framework or as a standalone code. Extensive documentation is provided in \cite{Workgroup:2017lvb} and the code itself for either use case. Future plans for the code include the extension of \ddcalc to handle models with velocity dependent scattering cross sections, the extension of the code to include likelihoods for cosmic axions searches, the addition of charged cosmic ray likelihoods, and the continued updating of our current library of likelihoods.

\ack I thank my fellow \gambit collaboration members for five years of enlightening and fruitful collaboration, and the NSERC for its financial support of this work.

\section*{References}
\bibliography{TAUPbib}

\end{document}